\documentclass{aastex631}
\usepackage{upgreek}
\usepackage{amsmath, amstext}
\usepackage{array}
\usepackage{empheq}
\usepackage{comment}
\usepackage{mathrsfs}
\usepackage{textcomp}
\usepackage{enumitem}   
\usepackage{gensymb}
\usepackage{hyperref}
\usepackage{graphicx}
\usepackage[caption=false]{subfig}
\usepackage{multirow}
\usepackage{float}
\usepackage{longtable}
\usepackage{needspace}
\usepackage{tabularx}
\usepackage[flushleft]{threeparttable}
\usepackage{booktabs} 
\usepackage{CJKfntef,CJK}
\usepackage{rotating}
\usepackage{booktabs}

\hypersetup{colorlinks, linkcolor={blue}, citecolor={blue}, urlcolor={blue}}

\definecolor{DarkOrange}{RGB}{204, 85, 0}
\definecolor{LincolnGreen}{RGB}{17, 102, 0}

\usepackage{xspace}

\newcommand{\cxo }{{\it Chandra}}

\newcommand{\vla }{{\it VLA}}
\newcommand{\atca }{{\it ATCA}}
\newcommand{\hst }{{\it HST}}

\begin{document}
\begin{CJK*}{UTF8}{gbsn}
\pagenumbering{arabic}

\title{Searching for Black Hole Candidates in Quiescence by Using Multi-band Observations in Globular Cluster M22 (NGC 6656)}
\correspondingauthor{Wei-Min Gu, Shan-Shan Weng}
\email{guwm@xmu.edu.cn, wengss@njnu.edu.cn }

\author[0009-0000-3830-9650]{Yu-Jing Xu (徐雨婧)}
\affiliation{Department of Astronomy, Xiamen University, Xiamen, Fujian 361005, People's Republic of China}

\author[0000-0003-3137-1851]{Wei-Min Gu (顾为民)}
\affiliation{Department of Astronomy, Xiamen University, Xiamen, Fujian 361005, People's Republic of China}

\author[0000-0001-9321-6000]{Yuanqi Liu (刘媛琪)}
\affiliation{Shanghai Astronomical Observatory, 80 Nandan Road, Shanghai 200030, People's Republic of China}

\author[0009-0001-6886-8397]{Xin-Yu Fang (方欣雨)}
\affiliation{Department of Astronomy, Xiamen University, Xiamen, Fujian 361005, People's Republic of China}

\author[0000-0001-7595-1458]{Shan-Shan Weng (翁山杉)}
\affiliation{School of Physics and Technology, Nanjing Normal University, Nanjing, Jiangsu 210023, People's Republic of China}

\author[0000-0003-4341-0029]{Tao An (安涛)}
\affiliation{Shanghai Astronomical Observatory, 80 Nandan Road, Shanghai 200030, People's Republic of China}

\begin{abstract}

We present a multi-wavelength investigation of radio sources in the globular cluster M22 (NGC 6656) using the \textit{Karl G. Jansky Very Large Array}, \textit{Chandra} and \textit{Hubble Space Telescope}. By cross-matching the radio and X-ray source catalogs, we identify eight radio/X-ray counterparts, of which VLA22 is the most promising stellar-mass black hole (BH) candidate. Its radio and X-ray luminosities are consistent with the established $L_{\rm X}-L_{\rm R}$ correlation for BH low-mass X-ray binaries. The observed X-ray variability supports an accreting nature.
One possible optical counterpart ($M_{\mathrm{814}} \approx 16.695 \pm 0.011$ mag) is identified. Based on its inferred stellar parameters, the estimated orbital period of $P_{\rm orb} \sim 16 \pm 6~{\rm h}$ places the proposed counterpart predominantly in the BH region rather than in the NS region in the $L_{\rm X}-P_{\rm orb}$ plane. These results demonstrate the effectiveness of joint radio, X-ray, and optical observations in identifying quiescent BH candidates in globular clusters.

\end{abstract}

\keywords{globular clusters (656): individual (M22) - stars: black holes (162) - X-ray binaries (1811)}

\vspace{1em}

\section{Introduction}
Globular clusters (GCs) are highly efficient environments for the formation of exotic stellar systems. Despite accounting for only $\sim 0.1\%$ of the Galactic stellar mass, they host approximately $10\%$ of the low-mass X-ray binary (LMXB) population \citep{Clark1975}.  Historically, theoretical studies predicted that stellar-mass BHs should be efficiently ejected from GCs. Mass segregation drives BHs toward the cluster center, where repeated three-body interactions impart sufficiently large recoil velocities to eject most BHs from the cluster leaving at most one or no BHs in present-day clusters \citep{Kulkarni1993, Sigurdsson1993}. However, this paradigm has been challenged by the detection of super-Eddington sources in extragalactic GCs \citep{Maccarone2007, Irwin2010} and recent direct $N$-body simulations, which suggest that a significant BH population can survive and remain dynamically active over Hubble timescales \citep{Sippel2013, Morscher2013}. As ancient stellar systems with ages exceeding 12 Gyr, GCs have produced large populations of compact remnants, including white dwarfs (WDs), neutron stars (NSs), and black holes (BHs) \citep{Heger2003}.

Identifying BH candidates in GCs is challenging because most reside in a quiescent state with low accretion rates, making them difficult to distinguish from NS-LMXBs using X-ray data alone. The widespread deployment and continued improvement of high-sensitivity radio interferometers, most notably the Karl G. Jansky Very Large Array \citep[\vla{}, ][]{PerleyEVLA2011} and the Australia Telescope Compact Array \citep[\atca{}, ][]{WilsonATCA2011}, have enhanced our ability to study  BH-LMXBs. BHs in quiescence or in the hard state are characterized by a flat-to-inverted radio spectrum ($\alpha \gtrsim 0$; $\rm S \propto \nu^\alpha$) and are known to follow the empirical $L_{\rm X}-L_{\rm R}$ correlation, exhibiting systematically more radio luminous than NS-LMXBs at comparable X-ray levels \citep{Migliari2006}. Other radio-emitting compact objects commonly found in GCs exhibit different radio properties. Millisecond pulsars (MSPs) typically show steeper radio spectra \citep[$\alpha \approx -1.6$; ][]{Jankowski2018}, while cataclysmic variables (CVs) and active binaries (ABs) generally remain below detection limits at GC distances \citep{Drake1989, Abada-Simon1993, Osten2000, Marsh2016Natur, Russell2016}.

M22 (NGC 6656), located at $\sim 3.2$ kpc, is a prime target for such studies. The cluster has a foreground reddening of $E(B-V)\approx0.34$. Its structural parameters include a core radius ($r_{\rm c}$) of approximately $1.42'$ and a half-light radius ($r_{\rm h}$) of about $3.26'$ \cite{HarrisDist1996}. Previous X-ray and optical surveys identified numerous low-luminosity X-ray sources, including confirmed CVs \citep{webbXraySourcesTheir2004, webbCV1GlobularCluster2013}. Notably, \cite{straderTwoStellarmassBlack2012} identified two BH-LMXB candidates in M22 based on their flat radio spectra. However, these sources lacked X-ray counterparts, preventing a detailed investigation of their accretion properties. Therefore, combining radio, X-ray, and optical observations provides a more reliable means of identifying BH-LMXBs.

In this paper, we present a systematic multiwavelength study of M22 using observations from the \vla{}, \cxo{}, and \hst{}, with the goal of identifying and characterizing compact binaries within the cluster. Section 2 describes the data reduction and spatial matching; Section 3 presents the observational results and proposes a new BH-LMXB candidate; and Section 4 discusses the implications of our findings.

\section{Data access and processing}
        
\subsection{Radio observations} \label{sec:radio-obs}

The radio data used in this work were obtained from the Milky Way \atca{} and \vla{} Exploration of Radio sources In Clusters (MAVERIC) survey. This survey targeted 61 Galactic GCs using the \vla{} and \atca{}, with their high sensitivity and sub-arcsecond/arcsecond spatial resolution to identify compact sources \citep{shishkovskyMAVERICSurveyRadio2020a, tudorMAVERICSurveyCatalogue2022}. M22 was observed by the \vla{} in May 2011 (Project Code: 10C-109) with a total observing time of 10 h (7.3 h on-source). These observations used the C-band (4--8 GHz) with two 2 GHz bandwidths centered at 5.0 and 6.8 GHz (Table \ref{tab:radio_obs_log}). We adopted the catalog produced by \cite{shishkovskyMAVERICSurveyRadio2020a} and published via the VizieR platform\footnote{\protect{https://vizier.cds.unistra.fr/viz-bin/VizieR?-source=J/ApJ/903/73}} for our analysis, retaining the original source numbering based on flux density. For each source, we took the larger of the cataloged positional uncertainty and a systematic uncertainty corresponding to approximately 10\% of the synthesized beam size, \citep{shishkovskyMAVERICSurveyRadio2020a}, typically resulting in a final uncertainty of $0.15'' \times 0.1''$. The absolute astrometry of the VLA data was tied to the ICRF frame through phase referencing using nearby calibrator sources with accurately known positions, thereby ensuring accurate absolute astrometry. 

To investigate the radio variability of VLA22, we retrieved the raw data and reprocessed the four individual sub-observations that constitute the total program. Data calibration and imaging were performed using Common Astronomy Software Applications \citep[CASA; version 6.4.1.12][]{CASA2022}. The archival observations used the standard source 3C286 as the flux and bandpass calibrator and J1820-2528 as the phase calibrator. We imaged the data with Briggs weighting and applied primary beam correction to obtain reliable flux densities for sources located away from the phase center. For each epoch, we used the {\tt AEGEAN} source-finding algorithm \citep{HancockAgean2012} to identify sources and measure their flux densities. Only sources detected at $> 5\sigma$ significance were considered for further analysis. Notably, due to the shortened integration time per epoch, the rms noise level in each sub-observation is approximately twice that of the combined deep image. Consequently, many faint sources included in the deep MAVERIC catalog fall below the detection threshold in the individual epochs.

A similar data reduction procedure was applied to the observation in May 2014 (Project Code: SE0068), which was conducted at central frequencies of 5.0 and 7.4 GHz. These data were used to provide quasi-simultaneous radio coverage matching the X-ray observation. All the observational data used are listed in Table \ref{tab:radio_obs_log}.

\subsection{X-ray counterpart searching}

We used archival \cxo{} observations of M22 obtained in 2005 (16 ks; ObsID: 5437) and 2014 (86 ks; ObsID: 14609) using the Advanced CCD Imaging Spectrometer (ACIS). While \cite{bahramianMAVERICSurveyChandra2020} provided a high-precision source catalog\footnote{\protect{http://vizier.cds.unistra.fr/viz-bin/VizieR?-source=J/ApJ/901/57}} (GCcat) for 38 GCs, we compared this catalog with the \cxo{} Source Catalog Release 2.0 \citep[CSC 2.0, ][]{Evans2010}. Due to its superior astrometric calibration against the Gaia DR3 reference frame, we adopted CSC 2.0 positions for spatial matching, while X-ray luminosities at $1\text{--}10$ keV were derived from GCcat. We extracted all X-ray sources within a $5'$ radius centered on M22. Positional uncertainties were conservatively defined as the larger of the $2\sigma$ cataloged error and a typical on-axis value\footnote{\protect{https://cxc.harvard.edu/csc2/about.html}} of $0.5''$ \citep{GarmireCXO2003}. The Chandra observations used in this work are available in the Chandra Data Collection (doi: {\dataset[DOI: 10.25574/cdc.593]{https://doi.org/10.25574/cdc.593}}).

Initial cross-matching between the \vla{} and \cxo{} catalogs was performed using the Python package {\tt Astropy}. We considered two sources to be associated when their 2$\sigma$ positional uncertainty regions overlapped. Specifically, a radio and X-ray source pair was considered a potential match if their angular separation satisfied
\begin{equation}
\Delta {\rm pos} = \sqrt{(RA_{\rm X} - RA_{\rm R})^{2}cos^{2}(DEC_{\rm X})+(DEC_{\rm X} - DEC_{\rm R})^{2}} < R_{\rm X} + R_{\rm R},
\end{equation}
where $RA$ and $DEC$ denote the right ascension and declination of the sources, respectively. $R_{\rm X}$ and $R_{\rm R}$ represent the positional uncertainties discussed above. Under this criterion, an X-ray source was identified as the counterpart of a radio source when the above condition was satisfied.
This procedure identified eight radio/X-ray counterparts listed in Table \ref{tab:cat}. For detailed spectral analysis, raw data were reprocessed using the \cxo{} Interactive Analysis of Observations\footnote{\protect{http://cxc.harvard.edu/ciao/}} ({\tt CIAO}) software \citep[][ CALDB: 4.17]{FruscioneCIAO2006} via the ${\tt chandra\_repro}$ script. Spectra were extracted using ${\tt specextract}$, and spectral modeling was conducted in ${\tt XSPEC}$ (v12.15.0) \citep{ArnaudXspec1996}.

\subsection{Hubble optical observations} \label{sec:match}

The \hst{} UV Globular Cluster Survey \citep[HUGS: {\dataset[10.17909/T9810F]{http://dx.doi.org/10.17909/T9810F}};][]{https://doi.org/10.17909/t9810f,PiottoHUGs2015, NardielloHUGS2018} provides a high-precision photometry for M22, including \hst{}/WFC3 in F275W ($\mathrm{UV_{275}}$), F336W ($\mathrm{U_{336}}$), and F438W ($\mathrm{B_{438}}$), as well as \hst{}/ACS in F606W ($\mathrm{V_{606}}$) and F814W ($\mathrm{I_{814}}$). We adopted HUGS as our primary dataset for constructing Color-Magnitude Diagram (CMD), and further integrated the WFPC2/F656N ($\mathrm{H}\alpha$) narrow-band data (Obs ID: 5344) into our CMDs to facilitate the study of $\mathrm{H}\alpha$ emission/absorption features. To investigate variability, we integrated 62 observations of archival \hst{}/WFPC2 $\mathrm{I_{814}}$ data from 1999 (ObsID: 7615). Despite the two-decade gap, photometric consistency was verified using non-variable stars. Data were retrieved from MAST{\footnote{\protect{https://mast.stsci.edu/search/ui/\#/hst}}} \citep[doi: {\dataset[10.17909/8j9r-yr64]{http://dx.doi.org/10.17909/8j9r-yr64}};][]{https://doi.org/10.17909/8j9r-yr64}, and PSF photometry was performed using {\tt DOLPHOT} \citep{Dolphin2016}, retaining only fits with $\chi^{2} < 5$. Magnitude values are reported without extinction correction unless specified.

Astrometric matching was prioritized using ACS images, because the sharper PSF and lower level of scattered light provide better resolution in crowded fields \citep{SirianniACSWFPC2005}. Furthermore, the ACS solution is precisely anchored to the Gaia frame, whereas WFC3 and WFPC2 images exhibit slight frame drifts. Consequently, we first identified optical counterparts in the ACS frames before extracting multi-band photometry. We note that several faint counterparts detected in ACS remain below the detection threshold in the older WFPC2 data, precluding variability analysis for those specific targets.

\begin{figure*}
    \centering
    \makebox[\textwidth][c]{\includegraphics[width=1\linewidth]{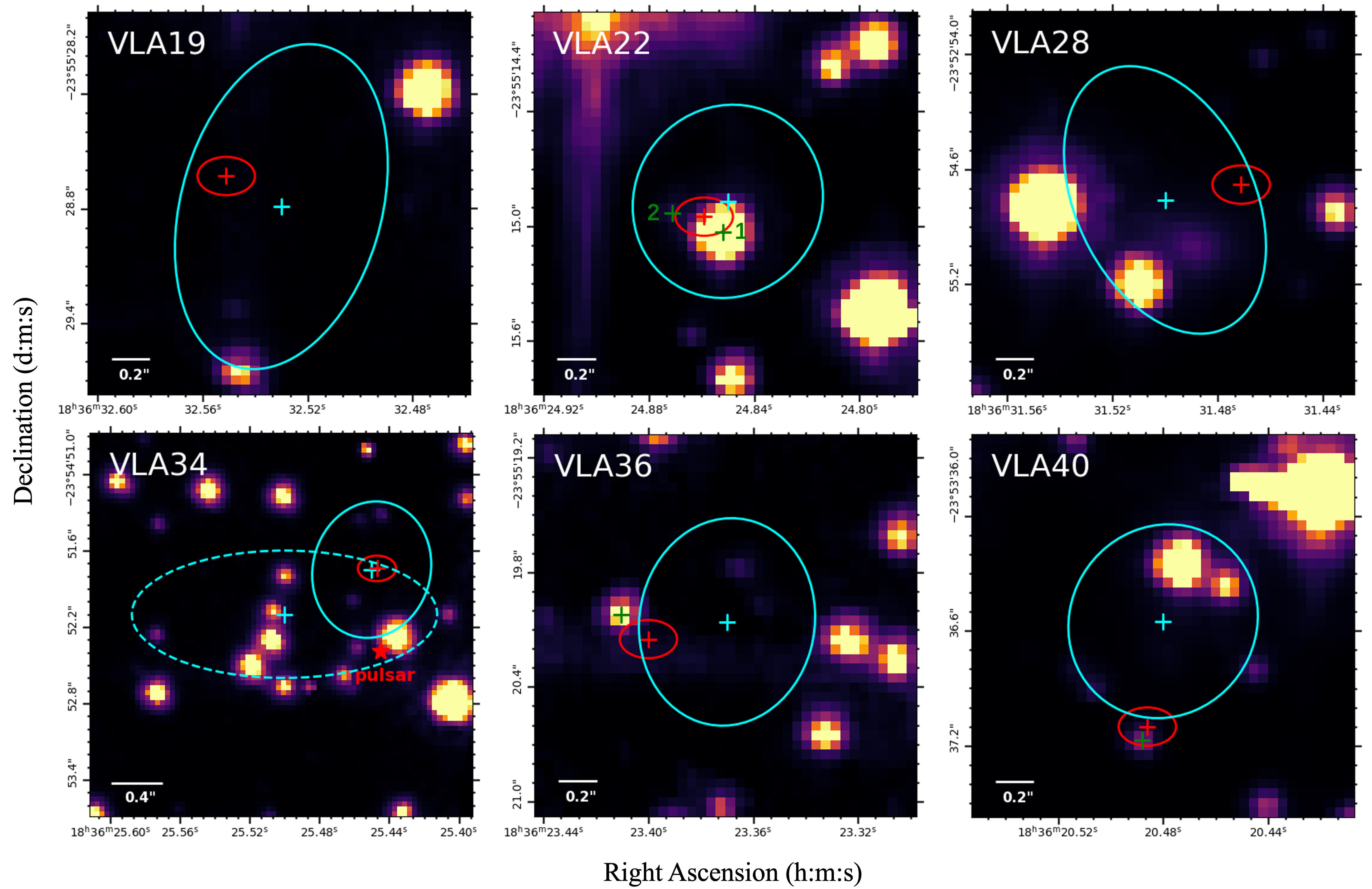}}
    \caption{Multiwavelength identification images of six sources observed in the radio, optical and X-ray bands. The color background map shows the corresponding optical/infrared background from \hst{}. For VLA19, VLA22, VLA28, VLA36 and VLA40, the optical charts are 1.5$''$ $\times$ 1.5$''$ in size obtained with  ACS/F814W. For VLA34, the optical charts are 3$''$ $\times$ 3$''$ in size given by WFC3/F814W. The cyan cross and ellipse indicate the position and 95\% error range of the coordinate given by \cxo{}, while the small red cross and ellipse indicate the radio position and positional uncertainty corresponding to 10\% of the synthesized beam size (0.15$''$ $\times$ 0.1$''$). The green cross marks the possible optical counterpart. The two possible optical counterparts of VLA22, namely VLA22-1 and VLA22-2, are labeled with the identification numbers 1 and 2, respectively. The red star marks the position of a known pulsar PSR J1836-2354A (M22A), and the dotted ellipse outlines the calibrated X-ray counterpart position to the pulsar given by \cite{amatoPulsarEmission2019}. The scales are also given at the lower left corner of each panel.}
    \label{fig:img}
\end{figure*}

Among the 8 identified X-ray/radio counterparts, six (VLA19, VLA22, VLA28, VLA34, VLA36, and VLA40) lie within the \hst{} field of view. Their radio and X-ray positions, including uncertainty ellipses, are overlaid on the \hst{}/ACS imaging in Figure \ref{fig:img}.

\begin{deluxetable*}{lcccccccc}
\tablecaption{Log of VLA Radio Observations for M22 \label{tab:radio_obs_log}}
\tablehead{
\colhead{Project} & \colhead{Date} & \colhead{Band} & \colhead{Config.} & \colhead{Time} & \colhead{Beam$_{\rm low}^{a}$} & \colhead{$\sigma_{\text{RMS, low}}^{ab}$} & \colhead{Beam$_{\rm high}^{a}$} & \colhead{$\sigma_{\text{RMS, high}}^{ab}$} \\
\colhead{} & \colhead{} & \colhead{} & \colhead{} & \colhead{(h)} & \colhead{($\text{arcsec} \times \text{arcsec}$)} & \colhead{($\mu$Jy/beam)} & \colhead{($\text{arcsec} \times \text{arcsec}$)} & \colhead{($\mu$Jy/beam)}
}
\startdata
10C-109 & Combined & C & B/BnA & 7.3 & $1.54 \times 0.81$ & 2.4 & $1.14 \times 0.59$ & 2.0 \\
 & 2011-05-22  & C & BnA & 1.8 & $3.00 \times 0.34$ & 4.6 $\sim$ 7.3 & $0.91 \times 0.26$ & 4.3 $\sim$ 10.4 \\
 & 2011-05-23  & C & BnA & 1.8 & $1.77 \times 0.34$ & 4.5 $\sim$ 7.6 & $1.39 \times 0.27$ & 4.3 $\sim$ 10.4 \\
 & 2011-05-25  & C & B & 1.8 & $1.89 \times 0.36$ & 4.2 $\sim$ 7.4 & $1.46 \times 0.27$ & 3.9 $\sim$ 10.0 \\
 & 2011-05-29  & C & B & 1.8 & $1.20 \times 0.35$ & 4.9 $\sim$ 7.6 & $0.91 \times 0.25$ & 4.2 $\sim$ 11.0 \\
\hline
SE0068 & May 2014 & C & A & 2.24 & $0.31 \times 0.16$ & 3.5 $\sim$ 5.3 & $0.22 \times 0.11$ & 2.8 $\sim$ 8.5 \\
\enddata
\tablenotetext{a}{``low" refers to 5.0~GHz for both observations; ``high" refers to 6.8~GHz for Project 10C-109 and 7.4~GHz for the second Project SE0068.}
\tablenotetext{b}{The rms range is derived from the region between the center of M22 and $r_h$, due to primary beam correction.}
\end{deluxetable*}

\begin{deluxetable*}{lcccccccccl}
\tabletypesize{\footnotesize}
\tablecaption{Radio properties of cross-matched sources in M22 \label{tab:cat}}
\tablehead{
\colhead{VLA} & \colhead{$\alpha_{\rm VLA}$} & \colhead{$\delta_{\rm VLA}$} & \colhead{$S_{5\,\mathrm{GHz}}$} & \colhead{$S_{6.8\,\mathrm{GHz}}$} & \colhead{Rad.$^{a}$} & \colhead{loc.$^{b}$} & \colhead{$\alpha$$^{c}$} & \colhead{2CXO} & \colhead{$F_{\rm 1-10~keV}$} & \colhead{$\Delta$pos$^{d}$} \\
\colhead{} & \colhead{(h:m:s)} & \colhead{($\degree$:$'$:$''$)} & \colhead{($\mu$Jy)} & \colhead{($\mu$Jy)} & \colhead{($'$)} & \colhead{} & \colhead{} & \colhead{} & \colhead{(erg\,cm$^{-2}$\,s$^{-1}$)} & \colhead{($''$)}
}
\startdata
10 & 18:36:36.471 & $-$23:55:01.07 & 52.5 $\pm$ 2.8 & 40.4 $\pm$ 3.6 & 2.96 & $r_{\rm h}$ & $-$0.87 $\pm$ 0.35 & J183636.4$-$235501 & $6.11\times10^{-15}$ & 0.34 \\
19 & 18:36:32.551 & $-$23:55:28.63 & 26.0 $\pm$ 2.6 & 33.3 $\pm$ 2.5 & 2.3 & $r_{\rm h}$ & 0.79 $\pm$ 0.39 & J183632.5$-$235528 & $9.57\times10^{-16}$ & 0.33 \\
22 & 18:36:24.859 & $-$23:55:14.95 & 25.0 $\pm$ 2.7 & 19.5 $\pm$ 2.1 & 0.99 & $r_{\rm c}$ & $-$0.81 $\pm$ 0.51 & J183624.8$-$235514 & $6.62\times10^{-14}$ & 0.15 \\
25 & 18:36:09.529 & $-$23:53:21.18 & 24.1 $\pm$ 3.4 & 14.6 $\pm$ 3.9 & 3.42 & $...$ & $-$1.69 $\pm$ 0.97 & J183609.5$-$235320 & $1.66\times10^{-15}$ & 0.69 \\
28 & 18:36:31.471 & $-$23:52:54.68 & 23.3 $\pm$ 2.5 & 16.2 $\pm$ 2.6 & 2.2 & $r_{\rm h}$ & $-$1.23 $\pm$ 0.67 & J183631.5$-$235254 & $7.45\times10^{-16}$ & 0.41 \\
34 & 18:36:25.447 & $-$23:54:51.74 & 17.5 $\pm$ 2.8 & 11.9 $\pm$ 2.0 & 0.67 & $r_{\rm c}$ & $-$1.27 $\pm$ 0.80 & J183625.4$-$235451 & $1.28\times10^{-15}$ & 0.04 \\
36 & 18:36:23.400 & $-$23:55:20.15 & 16.4 $\pm$ 2.6 & 8.7 $\pm$ 2.2 & 1.06 & $r_{\rm c}$ & $-$2.03 $\pm$ 0.85 & J183623.3$-$235520 & $1.44\times10^{-15}$ & 0.42 \\
40 & 18:36:20.486 & $-$23:53:37.10 & 14.7 $\pm$ 2.6 & $<$6.4 $\pm$ 2.4 & 1.03 & $r_{\rm c}$ & $<$0.4 & J183620.4$-$235336 & $3.36\times10^{-14}$ & 0.56 \\
\enddata
\tablenotetext{a}{Rad. is the projected distance from the cluster center.}
\tablenotetext{b}{loc. denotes whether the source locates within the core radius ($r_{\rm c}$) or half-light radius ($r_{\rm h}$).}
\tablenotetext{c}{The radio spectral index $\alpha$, for $\rm S \propto \nu^\alpha$.}
\tablenotetext{d}{$\Delta$pos is the radio--X-ray positional offset.}
\end{deluxetable*}

\begin{figure*}
    \centering
    \includegraphics[width=0.7\linewidth]{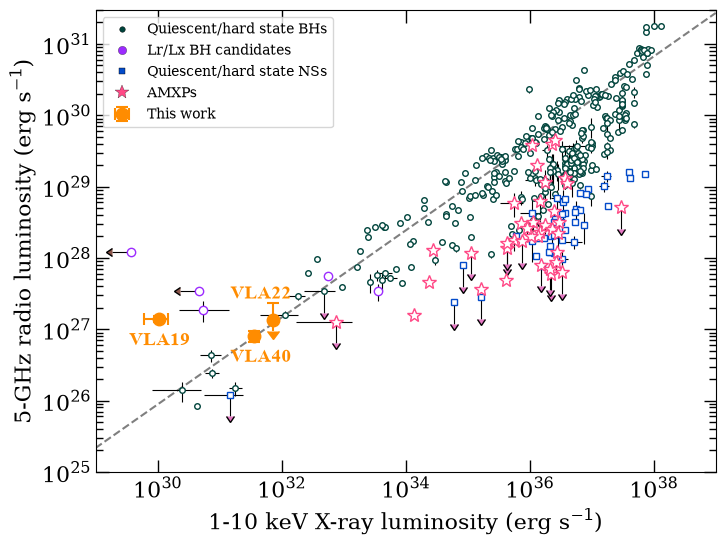}
    \caption{Radio luminosity at 5 GHz versus 1-10 keV X-ray luminosity. The comparison data and plotting framework were adapted from \citep{bahramian2022} and reference as well \citep{Tetarenko2016}. Green circles denote confirmed BH binaries, and blue squares denote NS binaries in quiescence or hard state. The purple dots are BH candidates selected by radio observation. The pink stars are accreting millisecond X-ray pulsars (AMXPs). The grey dashed line is the $L_{\rm X}-L_{\rm R}$ correlation of BH-LMXBs \citep{Gallolr2014}. The orange dots are sources we found in this work.}
    \label{fig:lr}
\end{figure*}

\section{Results}

\subsection{M22-VLA22: A Promising Quiescent Stellar-Mass Black Hole Candidate} \label{sec:vla22}

\subsubsection{Radio properties}
M22-VLA22 was detected at both 5 GHz and 6.8 GHz with a significance exceeding $5\sigma$ in the 2011 VLA observations. Located within the cluster core, it exhibits a radio spectral index of $\alpha = -0.81 \pm 0.51$. Although the best-fit value suggests a moderately steep spectrum, its large uncertainty remains consistent with the flat-to-inverted radio spectra typically observed in quiescent BH-LMXBs. Assuming a flat spectrum and a cluster distance of 3.2 kpc, we derive a 5 GHz radio luminosity of $L_{\rm R} \approx 1.53 \times 10^{27}$ erg s$^{-1}$ ($L_{\rm R}=4 \pi d^2 \nu S_{\nu}$, where $\nu$ is the observing frequency). This luminosity is derived from the combined deep image including all four observing epochs. The flux densities obtained from the four individual observing epochs are compiled in Table \ref{tab:vla22_radio}. As discussed in Section \ref{sec:radio-obs}, due to the shortened integration time per epoch, the shorter integration time in each individual epoch resulted in higher rms noise levels, preventing detections in some epochs and leaving only upper limits. Although the observed radio flux density varies between the individual epochs, we interpret these variations as stochastic modulations of the compact jet driven by accretion instabilities. Such behavior is consistent with the characteristic ``burbling" behavior observed in quiescent BH-XRBs \citep{Miller2008, chomiukM622013}. The alternative scenario of CVs will be independently evaluated in Section \ref{sec:redback}. It was not detected in the 2014 radio observation, possibly because of flux-density fluctuations similar to those seen in 2011.

\begin{table}
\centering
\caption{Transposed Radio Flux Density Measurements of M22-VLA22}
\label{tab:vla22_radio}
\begin{tabular}{lcccccc}
\toprule
Band & 2011-Combined & 2011-05-22 & 2011-05-23 & 2011-05-25 & 2011-05-29 & 2014-05-23 \\
\midrule
$S_{5\,\rm GHz}$ ($\mu\rm Jy$) & 25.0 $\pm$ 2.7 & $<$ 24.5 & 31.3 $\pm$ 4.1 & $<$ 22.5 & 43.3 $\pm$ 4.2 & $<$ 18.5 \\
$S_{7\,\rm GHz}$ ($\mu\rm Jy$) & 19.5 $\pm$ 2.1 & $<$ 22.0   & $<$ 23.5 & 24.7 $\pm$ 4.2 & $<$ 24.0 & $<$ 16.5 \\
\bottomrule
\end{tabular}

\end{table}

\begin{figure*}
    \centering
    \includegraphics[width=1\linewidth]{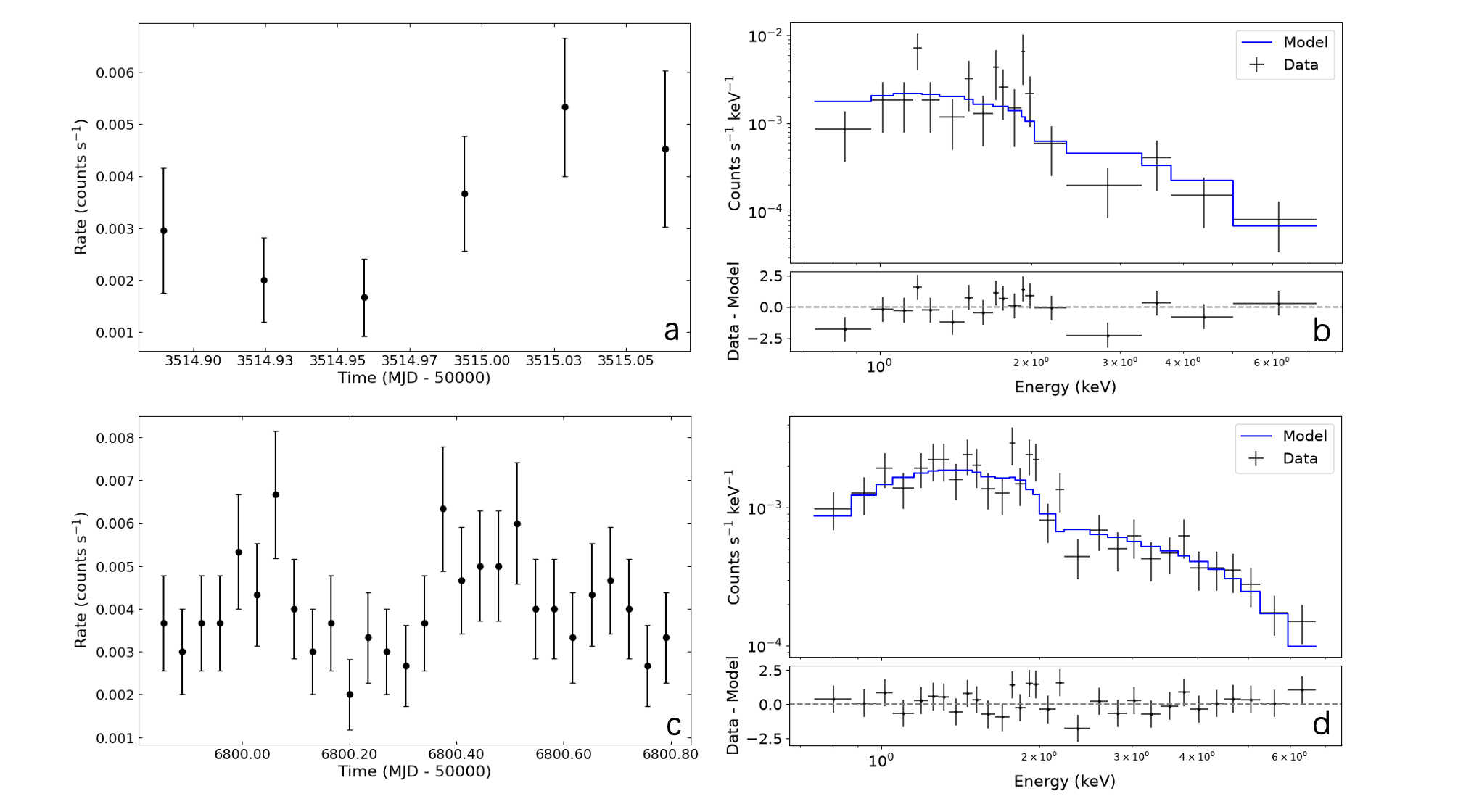}
    \caption{X-ray temporal and spectral characteristics of VLA22 observed with Chandra ACIS-S. Left: X-ray light curves from ObsID 5437 (panel a) and ObsID 14609 (panel c), displayed with a 3000 s time resolution. Right: The corresponding X-ray spectra for these observations are shown in panels b and d. In the spectral plots, the black points with error bars represent the observed data, while the solid blue curves indicate the best-fit results obtained from a single power-law model.}
    \label{fig:xray}
\end{figure*}

\begin{table}
    \centering
    \caption{X-ray Spectral Results for VLA22}
    \label{tab:vla22_xray}
    \begin{tabular}{lcccccc}
        \toprule
        ObsID & Year & Exposure & Photon Index  & $F_{\rm 0.5-7~keV}$ & $F_{\rm 1-10~keV}$   & $\chi^{2}$/dof \\
              &      & (ks)     & ($\Gamma$)   & (erg s$^{-1}$ cm$^{-2}$)      & (erg s$^{-1}$ cm$^{-2}$) \\
        \midrule
        5437 & 2005 & 16       & $1.57^{+0.43}_{-0.41}$ & $4.06 \times 10^{-14}$ & $4.93 \times 10^{-14}$  & 20.32/29 \\
        14609  & 2014 & 86       & $1.20_{-0.19}^{+0.19}$ & $5.05 \times 10^{-14}$ & $6.93 \times 10^{-14}$  & 22.24/16 \\
        \bottomrule
    \end{tabular}
\end{table}

\subsubsection{X-ray properties}
As shown in Figure \ref{fig:img}, VLA22 is spatially coincident with the X-ray source 2CXO J183624.8$-$235514, with a positional offset of only 0.15$''$, well within the 2$\sigma$ X-ray positional uncertainty. The GCcat reports a 1--10 keV X-ray luminosity of $L_{\rm X} = 7.20 \times 10^{31}$ erg s$^{-1}$. In the $L_{\rm X}-L_{\rm R}$ plane (Figure \ref{fig:lr}), VLA22 is consistent with the empirical BH track when plotted using the 2011 radio detection and the 2014 radio upper limit, combined with the 2014 X-ray luminosity. Although VLA22 was not detected in the radio band in 2014, the 2011 detection and the 2014 upper limit are both consistent with a quiescent state. Future simultaneous observations with concurrent detections in both bands remain essential for eliminating uncertainties associated with long-term variability.

The X-ray temporal and spectral properties are shown in Figure \ref{fig:xray} and Table \ref{tab:vla22_xray}. The light curve shows variability on timescales of a few hours (Figure \ref{fig:xray}), while the peak signal in the Lomb–Scargle power spectrum remains below the 2$\sigma$ confidence level. The short-term variability is consistent with stochastic accretion-driven changes rather than a stable periodic
modulation. Periodic X-ray modulation arising from orbital obscuration is uncommon among known BH-LMXBs, has been reported only in the low/hard state (e.g. Cyg X-1, \citealp{BrocksoppCyg1999}; 1E 1740.7-2942, and GRS 1758-258, \citealp{Smith2002}), whereas no such periodicities have been observed in quiescence. Nevertheless, long-term X-ray monitoring may reveal orbital periodicities in some accreting binaries \citep{Wen2006}. We modeled the spectra using an absorbed power-law model with the hydrogen column density fixed at the foreground value of M22 \citep[$1.97 \times 10^{21} \text{ cm}^{-2}$;][]{Cheng2018}. The resulting photon indices are $\Gamma = 1.57^{+0.43}_{-0.41}$ and $\Gamma = 1.20^{+0.19}_{-0.19}$ for the two observations in 2005 and 2014, respectively. These photon indices are harder than the typical value of $\Gamma \sim 2.1$ reported for most quiescent BH-LMXBs \citep{Plotkin2013}. 
The inferred photon index can be sensitive to the adopted $N_{\rm H}$. Only a few exceptional quiescent BH-LMXBs, such as GRO J1655-40 and XTE J1550-564, have been reported to exhibit relatively hard spectra ($\Gamma \sim 1.3$) when $N_{\rm H}$ was fixed to the value inferred from optical extinction \citep{Kong2002,Hameury2003,Corbel2006}, whereas softer spectra ($\Gamma \sim 2.1$) were obtained when $N_{\rm H}$ was allowed to vary freely \citep{Plotkin2013}. On the other hand, V4641 Sgr exhibits a notably hard quiescent X-ray spectrum ($\Gamma \sim 0.9$), which may be associated with jet emission \citep{Tomsick2003, Gallo2014}. These examples suggest that the relatively hard spectrum of VLA22 may have a more complex origin.
The observed X-ray spectrum is also notably harder than the typical spectra of quiescent NS-LMXBs \citep{Barret2001} and transitional MSPs \citep{Bogdanov2015}, which often feature a soft thermal component from the neutron star surface. Because hard, non-thermal X-ray spectra can also arise from intrabinary shocks in spider pulsar systems, we separately evaluate the redback MSP scenario in Section \ref{sec:redback}.

\subsubsection{Optical counterparts}

\begin{figure*}
    \centering
    \includegraphics[width=1\linewidth]{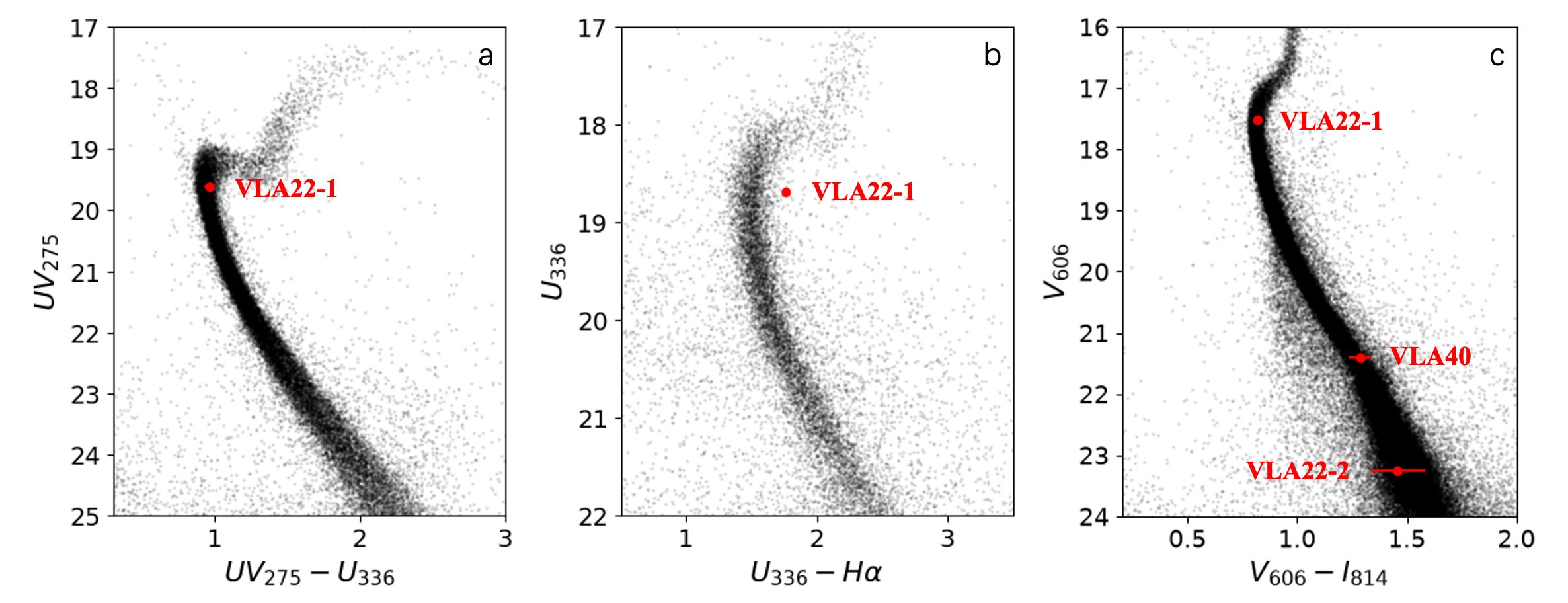}
    \caption{Color–magnitude diagram (CMD) of M22 following \cite{NardielloHUGS2018}. The red symbols denote the optical counterparts of VLA22 and VLA40, while black dots are other sources in M22. Data in panels a and c are obtained from the HUGS survey, whereas the data in panel b are from the HST/WFPC2 observations.}
    \label{fig:cmd}
\end{figure*}

\begin{figure}
    \centering
    \includegraphics[width=1\linewidth]{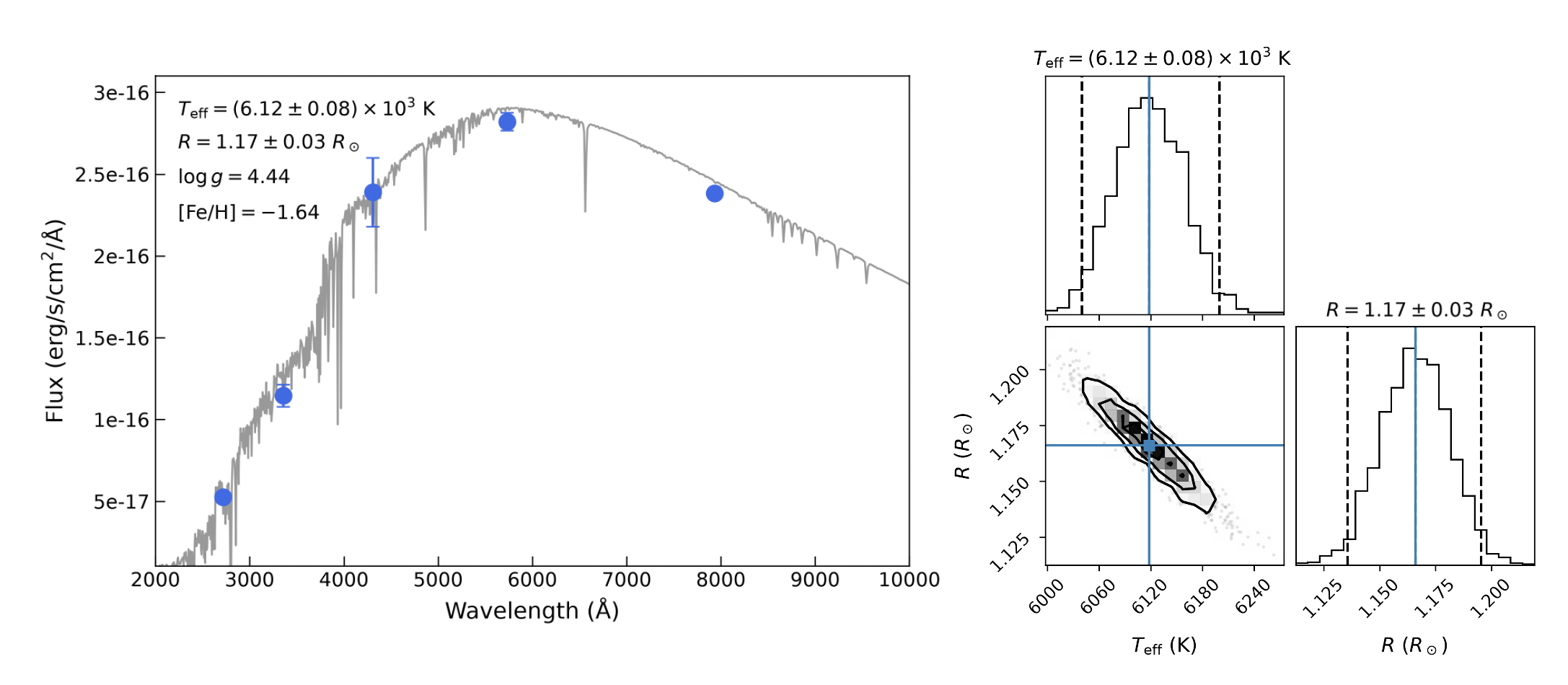}
    \caption{The spectral energy distribution (SED) fitting results for the optical counterpart VLA22-1. Left panel: The observed broadband photometric fluxes (blue points with $2\sigma$ error bars) overlaid with the best-fit synthetic stellar spectrum (grey solid line). During the fitting procedure, the visual extinction, surface gravity, and metallicity were fixed at $A_V = 1.05$, $\log g = 4.44$, and $\rm [Fe/H] = -1.64$, respectively. Right panel: The 1D and 2D posterior probability distributions for the effective temperature ($T_{\rm eff}$) and stellar radius ($R$) derived from the Markov Chain Monte Carlo (MCMC) sampling. The solid blue lines indicate the best-fit (median) values, while the vertical dashed black lines represent the $1\sigma$ confidence intervals.}
    \label{fig:sed}
\end{figure}

\begin{figure*}
    \centering
    \includegraphics[width=1\linewidth]{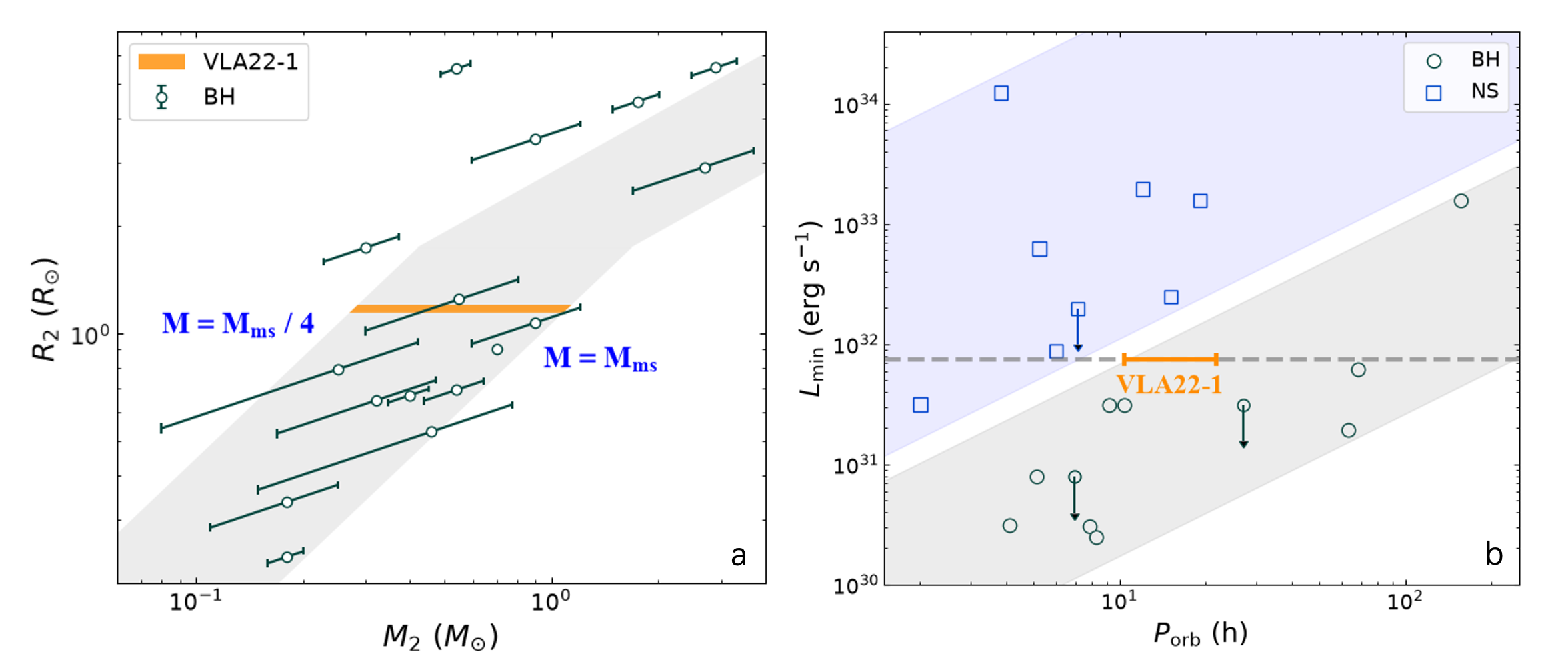}
    \caption{Panel a: Secondary mass ($M_2$) versus stellar radius ($R_2$) for confirmed BH-LMXBs. Open green circles and inclined lines represent $R_2$ derived from known $M_2$ and $P_{\rm orb}$ \citep{gelinoGROJ0422+32Lowest2003,Shao2020,Tetarenko2016ApJS} via the Roche-lobe filling constraint \citep[$R_2 \propto P_{\rm orb}^{2/3} M_2^{1/3}$;][]{Frank2002,Zheng2019}.
    The shaded gray region represents the bounds between standard MS ($M = M_{\rm ms}$) and stripped ($M = M_{\rm ms}/4$) donor stars, where $M_{\rm ms}$ is derived from the empirical mass-radius relation proposed by \cite{Demircan1991}. The orange region represents the range of $M_2$ and $P_{\rm orb}$ under the estimated $R_2 \approx 1.03 \pm 0.07~R_{\odot}$ of our target VLA22-1.
    Panel b: Orbital period versus X-ray luminosity of identified binary systems regarding BH event horizon proposed by \cite{NarayanBlack2008}. The hollow squares represent NSs and circles are BHs. The grey line is the X-ray luminosity level of VLA22 and the orange segment represents the estimated period of VLA22-1.}
    \label{fig:evt}
\end{figure*}

Within the \cxo{} error circle, we identified two potential optical counterparts: VLA22-1 ($M_{\mathrm{814}} \approx 16.695 \pm 0.011$ mag) and VLA22-2 ($M_{\mathrm{814}} \approx 21.805 \pm 0.086$ mag). 
However, morphological parameters of VLA22-2 reveal anomalously high RADXS values (ranging from 1.7 to 3.5 across both bands), which indicate severe profile distortion rather than a clean stellar point-source. These extreme values, combined with a membership probability of ${\rm PROB} = -1.0$, suggest that VLA22-2 is either an instrumental artifact contaminated by bright stellar diffraction spikes or a poorly constrained background object \citep{NardielloHUGS2018}. We therefore exclude VLA22-2 as a viable candidate. The following analysis therefore focuses on VLA22-1.

For VLA22-1, we extracted the $\mathrm{I_{814}}$ light curve from archival WFPC2 data. The raw light curve exhibits fluctuations of less than 0.1 mag, which are likely dominated by instrumental effects, including detector noise and spatial variations in the CCD response, rather than intrinsic variability. To correct for these systematics, we performed differential photometry using an ensemble of 59 neighboring reference stars (16 mag $< M_{\mathrm{814}} <$ 18 mag, $\Delta M_{\mathrm{814}} \leq 0.1$ mag). These comparison stars showed systematic variations similar to those of VLA22-1. After subtracting the mean ensemble variation, the calibrated light curve of VLA22-1 shows a reduced scatter of $\Delta M_{\mathrm{814}} \sim 0.04$ mag. No statistically significant variability is detected within our sensitivity limit, consistent with a binary viewed at low inclination. Otherwise, a tidally distorted donor star would produce distinct ellipsoidal modulations in the optical light curve, due to changing projected surface area over the orbital phase in a high inclination binary \citep{Morris1985}.
The CMD (Figure \ref{fig:cmd}) indicates that VLA22-1 exhibits a slight $\mathrm{H}\alpha$ excess, a classic tracer of an active accretion disk \citep{Charles2006}, while lacking the strong UV excess typically observed in CVs. Given that the optical luminosity is approximately two orders of magnitude greater than the X-ray luminosity, the optical emission is dominated by the donor star rather than the accretion disk. 
To estimate the contribution of the accretion disk to the optical continuum, we adopted the empirical optical–X-ray correlation for BHXBs in the hard and quiescent states from \cite{Russell2006} ($L_{\rm opt} = 10^{13.2 \pm 0.8} L_{\rm X}^{0.60 \pm 0.02}$). For the observed $L_{\rm X} \approx 6.62 \times 10^{31}~{\rm erg~s}^{-1}$, the relation predicts an optical disk luminosity of only $\sim 2.0 \times 10^{32}~{\rm erg~s}^{-1}$ in low/hard state, approximately 15 times lower than the stellar luminosity. The disk contribution is therefore neglected in the spectral energy distribution (SED) fitting. We used the ${\tt stellarSpecModel}$ package to generate synthetic stellar templates. During the fitting procedure, the visual extinction and metallicity were fixed at $A_V = 1.05$ and $\rm [Fe/H] = -1.64$ for M22 \citep{HarrisDist1996}, respectively. Given that broadband photometry is largely insensitive to surface gravity, we fixed the surface gravity at the solar value $\log g = 4.44$. Through $\chi^2$ minimization coupled with a Monte Carlo approach \citep{emcee}, we derived the best-fit parameters (Figure \ref{fig:sed}), yielding an estimated effective temperature of $T_{\rm eff} \approx (6.12 \pm 0.08) \times10^{3}~{\rm K}$ and a stellar radius of $R_2 \approx 1.17 \pm 0.03~R_{\odot}$. Because mass transfer generally makes donor stars undermassive compared to standard MS stars of the same radius, we constrained the donor mass of VLA22-1 by comparing it with the donor-star distribution of confirmed BH-LMXBs in the $M_2-R_2$ plane. As shown in Figure \ref{fig:evt}a, the donor radii $R_2$ are estimated from given $M_2$ and $P_{\rm orb}$ \citep{gelinoGROJ0422+32Lowest2003,Shao2020,Tetarenko2016ApJS}, under the condition that the donor star fills its Roche-lobe \citep{Frank2002,Zheng2019}:
\begin{equation}
P_{\rm orb} = 2\pi \left[ \frac{(R_2/0.462)^3} {G M_2} \right]^{1/2} \ .
\label{E1}
\end{equation}
For a given $R_2$, we adopt the MS mass ($M_{\rm ms}$) as an upper limit and $M_{\rm ms}/4$ as a lower limit for a donor star. This range encompasses the majority of the donors, as indicated by the grey region in Figure \ref{fig:evt}a. Here $M_{\rm ms}$ is derived from the empirical mass-radius relation proposed by \cite{Demircan1991} :
\begin{equation}
R_2/R_{\odot} \approx \begin{cases}
1.06~(M_2/M_{\odot})^{0.945},~M_2<1.66~M_{\odot} \\
1.33~(M_2/M_{\odot})^{0.555},~M_2>1.66~M_{\odot} \ .
\end{cases}
\label{E2}
\end{equation}
Thus the derived stellar radius $R_2$ for VLA22-1 (orange band in Figure \ref{fig:evt}a) implies a donor mass of $M_2 \approx 0.71 \pm 0.43~M_\odot$ and an orbital period of $P_{\rm orb} \sim 16 \pm 6$ h. Notably, Figure \ref{fig:evt}a shows that six confirmed BHs (XTE J1550-564, GS 2023+338, GX 339-4, GS 1354-64, GRO J1655-40, V4641 Sgr) lie below the $M = M_{\rm ms}/4$ boundary. The donor stars in these systems have evolved off the main sequence as giants or subgiants \citep{Orosz2011, Casares1994, Darias2008, Casares2009, Gonzlez2008, Sadakane2006}, in which scenario the required $P_{\rm orb}$ to maintain Roche-lobe overflow is even longer. \cite{NarayanBlack2008} demonstrated that at a given $P_{\rm orb}$, BH systems are significantly fainter in X-rays than NS systems due to the advection of energy across the event horizon. To make a comparison with their sample, we converted our X-ray luminosity to the 0.5--10 keV band. As shown in the $L_{\rm X}-P_{\rm orb}$ plane (Figure \ref{fig:evt}b), the combination of its relatively long period and low X-ray luminosity ($L_{\rm X} = 7.44 \times 10^{31}$ erg s$^{-1}$ at 0.5-10 keV), VLA22-1 is more consistent with the BH region.

Based on the available evidence, VLA22-1 is the most plausible donor-star candidate for this suspected BH-LMXB system. Future spectroscopic observations will be crucial to conclusively confirm this identification.

\subsubsection{CV, AB, Redback Pulsar and AGN Scenarios} \label{sec:redback}

\cite{RidderCV2023} conducted a systematic analysis of 35 cataclysmic variables (CVs) and found that only one source, V2400 Oph, reached a radio luminosity of $\sim10^{27}$ erg s$^{-1}$ at 5 GHz, and only during an active outburst. SS Cyg also exhibited a brief radio flare reaching a comparable luminosity ($\sim 1.4 \times 10^{27}$ erg s$^{-1}$ at 15.5 GHz) \citep{Mooley2017}. The radio variability of VLA22 can be compared with that observed during the 2010 outburst of SS Cyg. After reaching peak flux density, SS Cyg exhibited a secondary re-brightening during its decay phase but remained fainter than the initial peak \citep{RussellSSCyg2016}. In contrast, VLA22 exhibits the opposite behavior. For VLA22 to be a CV, this interpretation would require the system to undergo two distinct outbursts caught by chance within 6 days, as shown in Table \ref{tab:vla22_radio}. Such a coincidence is unlikely, and to our knowledge, no comparable case has been reported.

Furthermore, \cite{Guedel1993} established the empirical relation that $L_X/L_R \approx 10^{15}$ of chromospherically ABs. The X-ray emissions of ABs are intrinsically low ($L_{\rm X} < 10^{31} \text{ erg s}^{-1}$) and radio luminosities are well below detection limits at GC distances \citep{Drake1989}. This conclusion is supported by the work of \citet{Paduano2024}, who cross-matched 1,403 high-confidence, nearby active binaries from ZTF \citep{Chen2020} with the VLASS radio database and found zero radio detections. 

Given its non-thermal and hard X-ray spectrum, VLA22 shares several observational characteristics with redback pulsars, where non-thermal X-rays are driven by intrabinary shocks \citep{Roberts2015}. For example, the redback VLA38 in Terzan~5 lies near the empirical BH-LMXB track at low luminosities. Nevertheless, VLA38 features an exceptionally steep radio spectrum ($\alpha = -2.7_{-0.5}^{+0.7}$) \citep{urquhartMAVERICSurveyNew2020}, aligning with typical redbacks, whereas VLA22 is nominally much flatter. Furthermore, redback companions typically exhibit strong orbital modulations, whereas VLA22-1 shows no evidence of such modulation. 

To estimate the probability that VLA22 is an unrelated background AGN, we adopted the latest radio source-count statistics from \citet{Galluzzi2025}. Using the relation $N_{\rm bg}=\pi r^{2}N(>S)$, we integrated the radio source counts over the flux-density range of $10-60~\mu$Jy. Adopting the projected distance of VLA22 from the cluster center ($r=0.99'$) as the search radius, we estimate that approximately two background radio sources are expected within this region. Thus, source-count statistics alone do not provide strong evidence against a background-AGN interpretation.

Nevertheless, based the current multi-wavelength evidence, we are inclined to interpret VLA22 as a quiescent BH-LMXB, although its nature remains uncertain. Deeper multiband observations and high-cadence timing will be required to determine its nature.


\subsection{M22-VLA19, VLA40: stellar-mass BHs or Background Contaminants}

VLA19 lies at a projected distance of 2.3$'$ from the cluster center, between the core and the half-light radius. The radio source is associated with a faint X-ray counterpart ($L_{\rm X} \approx 7.79 \times 10^{30}$ erg s$^{-1}$) which exhibits a relatively soft X-ray spectrum ($\Gamma = 3.28 \pm 1.44$). Its radio spectrum is highly inverted ($\alpha = 0.79 \pm 0.39$), a feature strongly indicative of a self-absorbed compact jet typically associated with quiescent BH-LMXBs. Although the median radio spectral index of background AGNs is approximately $\alpha \sim -0.7$, their spectral-index distribution is broad, with rare sources exhibiting spectra ranging from $\alpha \sim -3$ to $\alpha \sim +1$ \citep{Huynh2015, Smol2017, shishkovskyMAVERICSurveyRadio2020a}. Nevertheless, its relatively large projected distance from the cluster center leaves open the possibility that it is a background AGN.

VLA40 was detected only at 5 GHz, with a $3\sigma$ upper limit at 6.8 GHz ($S_{\rm 6.8~GHz} < 6.4~\mu$Jy) corresponding to a spectral-index limit of $\alpha < 0.4$. It is spatially coincident (offset $\approx 0.56''$) with the X-ray source 2CXO J183620.4$-$235336, which exhibits a hard spectrum ($\Gamma = 1.54^{+0.35}_{-0.39}$) consistent with those of quiescent BHs. As shown in Figure \ref{fig:lr}, its position on the $L_{\rm X}-L_{\rm R}$ plane is consistent with the BH track. A potential optical counterpart ($M_{\mathrm{814}} \approx 20.213 \pm 0.031$ mag) was identified, showing stochastic variability of $\sim 0.15$ mag in the WFPC2 light curve. Such stochastic variability could arise from the reprocessing of fluctuating X-rays within the accretion disk or intrinsic turbulence in the accretion flow. However, we cannot rule out an extragalactic origin. The X-ray hardness and optical magnitude are also consistent with an AGN, in which the variability would instead originate from stochastic fluctuations in the accretion disk.

\subsection{M22-VLA34, VLA36: Millisecond Pulsar Candidates}

VLA34 is the source closest to the cluster core and is located only $0.04''$ from 2CXO J183625.4$-$235451. This X-ray source has previously been associated with the 3 ms pulsar PSR J1836$-$2354A (M22A). After applying the optical correction from \cite{amatoPulsarEmission2019}, both VLA34 and M22A remain within the $1\sigma$ X-ray error ellipse (Figure \ref{fig:img}). Based on the ATNF Pulsar Catalogue, M22A has a 1.4 GHz flux density of 0.2 mJy. Assuming a typical pulsar spectral index of  $\alpha = -1.6$ \citep{Jankowski2018}, the expected 5 GHz flux density is approximately $\approx 26~\mu$Jy, slightly higher than the observed flux of VLA34. However, VLA34 is offset from the timing position of M22A by $0.66''$. Given the high precision of both the interferometric radio positions and the pulsar timing solution, a physical association between VLA34 and M22A appears unlikely.

Although VLA34 and VLA36 lack detectable optical counterparts in the crowded core region, their radio spectral indices ($\alpha = -1.27 \pm 0.80$ and $-2.03 \pm 0.85$, respectively) are characteristically steep. Such steep spectra are consistent with the coherent radio emission from MSPs. The absence of optical counterparts is common for isolated rotation-powered pulsars, whose optical emission is typically well below the detection limits of current \hst{} observations at GC distances. However, radio spectral indices alone do not provide a definitive means of distinguishing MSPs from background AGN, as a small fraction of AGN can also exhibit similarly steep spectra. Additional multiwavelength observations and continued radio monitoring will be required to clarify their nature.

\subsection{M22-VLA10, VLA25, VLA28: Likely Extragalactic Background Sources}

VLA10 and VLA25 are located outside the {\it HST} field of view and both exhibit moderately steep radio spectra. VLA10 shows an exceptionally hard X-ray spectrum ($\Gamma = 0.65^{+0.46}_{-0.60}$), suggesting substantial photoelectric absorption of the soft X-ray emission, a characteristic commonly observed in obscured AGNs. 
VLA28 is covered by {\it HST} imaging but lacks a detectable optical counterpart. Its X-ray spectrum is consistent with the cluster environment ($N_{\rm H} \approx 2.82 \times 10^{21}~{\rm cm^{-2}}$; $\Gamma = 1.98^{+0.99}_{-1.4}$), but its steep radio index ($\alpha = -1.23 \pm 0.67$) is more consistent with either an MSP or a background AGN. Similarly, VLA25 lacks additional multi-wavelength constraints but its radio properties are consistent with a background source. Because all three sources (VLA10, VLA25, and VLA28) are located at large offsets from the core where the stellar density is low, they are statistically more likely to be extragalactic background sources rather than intrinsic cluster members. According to the GCcat, the expected number of background AGNs above the measured hard-band flux limit is 10.4, 8.0, and 11.9 within $r_{\rm h}$. These predicted background source densities further support the interpretation that the three sources are unrelated to M22.


\section{Conclusions and discussion} \label{sec:discuss}

In this study, we have revealed a mixed population of compact objects and background contaminants in M22. Among these, VLA22 emerges as a particularly promising BH-LMXB candidate based on the multi-wavelength analysis presented in Section \ref{sec:vla22}. If confirmed, it would represent the third such system identified in M22, following the previously reported VLA1 and VLA2 \citep{straderTwoStellarmassBlack2012}. Notably, unlike those earlier candidates, VLA22 has a clear X-ray counterpart and lies close to the established $L_{\rm X}-L_{\rm R}$ correlation for BH-LMXBs. The discovery of VLA22 reinforces the growing evidence that joint radio and X-ray diagnostics provide a powerful and effective approach to uncover quiescent BH systems in GCs, where traditional optical spectroscopy is often infeasible due to crowding and faintness. For the remaining sources, classification is limited by the lack of secure optical counterparts. In contrast, VLA34 and VLA36 are more naturally explained as MSPs, given their steep radio spectra and central locations within the cluster. The remaining sources (VLA10, VLA19, VLA25, VLA28 and VLA40) are most likely background sources.

Compared to VLA1 and VLA2, VLA22 lies farther from the cluster center. Following \cite{straderTwoStellarmassBlack2012}, one can estimate the expected mass segregation of stellar-mass BHs to obtain a rough
mass estimate as ($M_{\rm BH}/M_{\odot} = (r_{\rm core}/r_{\rm BH})^2$) \citep{Kulkarni1993}. Substituting the observed values of the core radius ($r_{\rm core}=1.33'$) and the projected BH distance ($r_{\rm BH}=0.99'$) into this relation yields an estimated BH mass of $\sim1.7~M_\odot$, which is substantially lower than the typical mass of stellar-mass BHs. Dense cluster cores are continuously subjected to dynamic multi-body interactions (e.g., BH-BH or BH-star encounters). These interactions frequently impart recoil kicks to BHs, which can scatter BHs into the cluster halo, after which they gradually sink back toward the core through dynamical friction on a timescale of $\sim10^7$ yr \citep{Sigurdsson1993}. Although the timescale is much shorter than the cluster age, N-body simulations show that BH ejections continue throughout the $\sim$12 Gyr evolution of globular clusters \citep{Morscher2013}. VLA22 may therefore have been observed during this phase of dynamical migration and the inferred mass of $\sim1.7~M_{\odot}$ should not be regarded as a robust mass estimate. This estimate neither rules out a neutron-star interpretation nor excludes the possibility that VLA22 hosts a stellar-mass BH.

\vspace{0.5cm}

\textit{Acknowledgements} -- 
We thank Zhiyuan Li and Yue Zhao for helpful discussion, and the referee for constructive suggestions that improved the paper. This work was supported by the National Key R\&D Program of China under grants 2023YFA1607900 and 2023YFA1607901, the National Natural Science Foundation of China under grants 12433007, 12473041 and 12221003. TA acknowledges support from the Shanghai Oriental Talent Project and Xinjiang Tianchi Talent Program. This research has made use of data collected by \vla{}, \cxo{} and \hst{}. 

\vspace{0.5cm}

\textit{Software} -- Astropy \citep{Astropy2022}, CIAO \citep{FruscioneCIAO2006}, DOLPHOT \citep{Dolphin2016}, Xspec \citep{ArnaudXspec1996}

\vspace{0.5cm}

\facilities{\vla{}, \cxo{}, \hst{}}

\bibliography{ref}
\bibliographystyle{aasjournal}



\end{CJK*}

\end{document}